\documentclass[a4paper]{article}

\usepackage[T2A]{fontenc}
\usepackage[cp1251]{inputenc}
\usepackage[russian]{babel}

\usepackage{graphicx}
\newcommand{\mmrus}{Математическое моделирование систем и процессов}

\begin{document}

\title{ВЕЙВЛЕТ-РЕГУЛЯРИЗАЦИЯ ОПЕРАЦИИ ДИФФЕРЕНЦИРОВАНИЯ СИГНАЛОВ С ШУМОМ}
\author{И.~А.~Патрикеев, Р.~А.~Степанов, П.~Г.~Фрик
\\Институт Механики Сплошных Сред, ул. Королева 1, 614013, г. Пермь}
\maketitle

%\begin{center}
%{\LARGE ВЕЙВЛЕТ-РЕГУЛЯРИЗАЦИЯ ОПЕРАЦИИ ДИФФЕРЕНЦИРОВАНИЯ СИГНАЛОВ С ШУМОМ }
%\\[5mm]
%{\large И.~А.~Патрикеев, Р.~А.~Степанов, П.~Г.~Фрик \\
%Институт Механики Сплошных Сред, ул. Королева 1, 614013, г. Пермь\\[5mm]
%28 декабря 2004 г.} \vspace{7mm}
%\end{center}

\begin{abstract}

Рассматриваются различные алгоритмы приближенного вычисления производной функции, заданной неточно. Алгоритм
дифференцирования с использованием вейвлет-преобразования сравнивается с алгоритмом на основе преобразования
Фурье и алгоритмами вычисления производной в физическом пространстве. Проведена численная оценка погрешности
вычислений различных алгоритмов на модельных примерах.

УДК 621.372\vspace{2mm}

\end{abstract}

{\bf Ключевые слова:} вейвлет, дифференцирование, фильтрация шума, регуляризация

\section{Введение}

Задача о численном дифференцировании функции, известной приближенно, является классическим примером некорректно
поставленной задачи, приводящей к неустойчивости решения \cite{Tikhonov}. Для обеспечения устойчивости по
Тихонову точное решение заменяют приближенным, которое управляется параметром регуляризации $\alpha$ и стремится
к точному при отсутствии погрешности измерений (подробнее о различных вариантах стабилизации решения см. в
\cite{Tikhonov,Pikalov,Troitsky}). На практике регуляризация обычно сводится либо к сглаживанию исходных данных
в физическом пространстве, либо к подавлению высоких частот в спектре измеренных данных. При этом оптимальная
ширина сглаживающего окна или соответствующая ему полоса пропускания фильтра связывается с ожидаемым уровнем
шума.

Представляется целесообразным сформулировать задачу регуляризации операции дифференцирования зашумленных данных
на языке вейвлет-представления сигналов, которое позволяет естественным образом совместить преимущества работы в
физическом пространстве и пространстве Фурье. Вейвлет-анализ, превратившийся за последнее десятилетие в хорошо
развитый раздел функционального анализа (см., например \cite{matias}), показал свою эффективность в задачах,
связанных с обработкой всевозможных многомасштабных сигналов и полей.

Первые попытки использования аппарата вейвлет-анализа при нахождении производной зашумленных данных были
выполнены в работах \cite{MM97,JBO99,my3,aa2002}. В данной работе методика вейвлет-дифференцирования описывается
в рамках общей проблемы без привязки к специфике сигнала. При этом проводится систематическое сравнение
вейвлет-регуляризации в задаче дифференцирования зашумленных сигналов с другими общеизвестными подходами.
Эффективность использования демонстрируется на конкретных примерах.

\section{Mетоды численного дифференцирования}

Пусть функция $f(x)$ имеет первую производную $g(x)$, так что

\begin{equation}
f(x)=\int_{0}^x g(x) dx
\label{funf}
\end{equation}

и определена на наборе точек $x_n$ с точностью до некоторой случайной ошибки $\xi$

\begin{equation}
\tilde{f}_n=f(x_n) + \xi. \label{funfi}
\end{equation}

Производная $g(x)$ выражается через $f(x)$

\begin{equation}
g(x)=\frac{\partial f(x)}{\partial x} \label{fung}
\end{equation}

и в простейшем случае может быть приближенно вычислена как

\begin{equation}
\tilde{g}_n=\frac{\tilde{f}_n - \tilde{f}_{n-1}}{\Delta x},
\label{fungi}
\end{equation}
где $\Delta x = x_n - x_{n-1}$.

В условиях шума формула (\ref{fungi}) может стать неустойчивой.
Принимая для случайной ошибки оценку $|\xi|\leq A$, можно записать
\begin{equation}
|g_n|\leq\left| \frac{f(x_n)-f(x_{n-1})}{\Delta x} \right| +
2A/\Delta x. \label{eq_abs_gi}
\end{equation}

При $\Delta x \to 0$ первое слагаемое в правой части (\ref{eq_abs_gi}) стремится к $g(x)$, а второе может быть
сколь угодно большим. Так, в частности, если функция задана на конечном интервале, то увеличение числа точек
приводит к увеличению второго слагаемого в (\ref{eq_abs_gi}).

\subsection{Дифференцирование в физическом пространстве}

В теории обобщенных функций \cite{Gelfand} вводится $\delta$-функция, задаваемая неявно как
\begin{equation}
\int_{-\infty}^{\infty} f(x) \delta(x) dx = \int_{0-}^{0+} f(x) \delta(x) dx= f(0),
\label{eq_delta}
\end{equation}
где $f(x)$ - гладкая функция.

Известно, что $\delta$-функция дифференцируема. Ее первая производная, обозначаемая
$\dot \delta$ и называемая "дублет", обладает свойством
\begin{equation}
g(x)=\int_{-\infty}^{\infty} f(x') \dot \delta(x-x') dx' = \int_{0-}^{0+} f(x-x') \dot \delta(x') dx'.
\label{eq_doublet}
\end{equation}

Отметим, что функция $\dot \delta$ нечетна и сосредоточена в
бесконечно малой окрестности точки $x=x'$. Функцию $\dot \delta$
называют иногда  {\it бесконечно малым диполем}.

\subsection{Дифференцирование в пространстве Фурье}

По теореме о производной \cite{Bracewell} дифференцирование в физическом пространстве сводится к умножению в
частотном пространстве. Фурье-образы функций $g$ и $f$ связаны соотношением
\begin{equation}
\hat{g}(k)=ik\hat{f}(k),
\label{eq_dif_ft}
\end{equation}
где $\hat{f}(k)$ - результат преобразования Фурье
\begin{equation}
\hat{f}(k)=\int_{-\infty}^\infty f(x) e^{-ikx} dx.
\label{eq_ft}
\end{equation}

Таким образом, находя Фурье-образ сигнала $\hat{f}(k)$, умножая
его в частотном пространстве на $ik$ и выполняя обратное
преобразование Фурье, можно найти производную $g(x)$:
\begin{equation}
g(x)=\frac{1}{2\pi}\int_{-\infty}^\infty ik \left( \int_{-\infty}^\infty f(x) e^{-ikx} dx \right)e^{ikx} dk.
\label{eq_ift}
\end{equation}

Отметим, что фильтр $ik$ имеет импульсную характеристику $\dot \delta$. То есть
\begin{equation}
ik=\int_{-\infty}^\infty \dot \delta (x) e^{-ikx} dx.
\label{eq_ik}
\end{equation}

Принципиальное отличие алгоритма (\ref{eq_ift}) от прямого
дифференцирования в физическом пространстве заключается в том, что
при вычислении преобразования Фурье используется информация о
сигнале во всех точках числовой оси, в то время как
дифференцирование является по определению операцией локальной.

\subsection{Дифференцирование с использованием
вейвлет-преобразования}

С точки зрения локальности, метод дифференцирования на основе
вейвлет-анализа занимает промежуточное положение между
дифференцированием в физическом пространстве и в пространстве
Фурье. Вейвлет-образ исходной функции $f(x)$ есть
 \begin{equation}
W_{a,b}\{f\} = \int \psi^*\left({{x-b}\over{a}}\right) f(x) dx,
\label{w_definition}
\end{equation}
 где в качестве анализирующего вейвлета $\psi(x)$ используется
комплексная или действительная функция, локализованная как в
физическом пространстве, так и в пространстве Фурье, а параметры
$a$ и $b$ определяют, соответственно, масштаб и положении функции
$\psi_{a,b}=\psi\left({{x-b}\over{a}}\right)$.

Записывая аналогичным образом вейвлет-образ функции $g(x)$ и
проводя дифференцирование по частям, легко получить
\begin{equation}
W_{a,b}\{g\} = \int_{-\infty}^\infty \psi_{a,b}^*(x) g(x) dx = -
\int_{-\infty}^\infty \frac{\partial\psi^*_{a,b}(x)}{\partial x}
f(x) dx \label{eq_wt}
\end{equation}

Таким образом, выполнив вейвлет-разложение функции $f(x)$ по
семейству $-\frac{\partial\psi_{a,b}(x)}{\partial x}$, и выполнив
затем обратное вейвлет-преобразование с помощью вейвлет-семейства
$\psi_{a,b}(x)$, можно получить искомую функцию $g(x)$:

\begin{equation}
g(x)= \frac{1}{C} \int_{0+}^\infty
\frac{da}{a^3}\int_{-\infty}^\infty db \psi_{a,x}(b) \left(
\int_{-\infty}^\infty dx \chi^*_{a,b}(x) f(x) \right), \label{iwt}
\end{equation}
где $\chi$ - анализирующий вейвлет, $\psi$ - синтезирующий
вейвлет, и $C$ - константа, определяемая выражением
\begin{equation}
C = {1\over {2\pi}} \int_{-\infty}^\infty
\frac{|\hat{\psi}(k)||\hat{\chi}(k)|}{|k|}dk ={1\over
{2\pi}}\int_{-\infty}^\infty |\hat{\psi}(k)|^2 dk < \infty.
\label{eq_C}
\end{equation}

Выбор пары $\chi$ и $\psi$ (для анализа и синтеза соответственно) из условия
\begin{equation}
\chi(x)= - \frac{\partial\psi(x)}{\partial x},
\label{eq_diffwav}
\end{equation}
предполагает, что производная от $\psi$ существует и является вейвлетом.

Например, пара функций $\chi=(1-x^2)\exp(-x^2/2)$ и $\psi=-x
\exp(-x^2/2)$ удовлетворяет условию (\ref{eq_diffwav}). В
принципе, любой вейвлет, имеющий первую производную, можно
использовать в качестве $\psi$. На практике, выбор конкретной пары
вейвлетов осуществляется с учетом специфики постановки задачи.

В качестве предельного случая можно использовать пару функций
$\chi=\dot{\delta}$ и $\psi=\delta$. Подставляя в уравнение
(\ref{iwt}) и рассматривая $\delta$-функцию, как сингулярный
вейвлет \cite{matias}, получаем

\begin{equation}
g(x)= \int_{-\infty}^\infty dx' g(x')\delta(x-x') = \int_{-\infty}^\infty dx' f(x') \dot{\delta}(x-x').
\label{eq_convdelt}
\end{equation}

\section{Регуляризация алгоритмов дифференцирования}\label{part_reg}

Рассмотренные выше алгоритмы не являются устойчивыми и для использования на практике требуют регуляризации.

\subsection{Приближенное вычисление производной в физическом
пространстве}

Для практического использования необходимо провести регуляризацию
уравнения свертки (\ref{eq_doublet}). Свертка с аппроксимирующей
дублет функцией, ширина которой управляется параметром
регуляризации $\alpha$, дает приближенное значение производной.
Обобщенная функция $\dot \delta$ аппроксимируется регулярной
нечетной функцией с нулевым средним значением, область локализации
которой определяется ожидаемым уровнем шума. Простейшей
аппроксимацией дублета на дискретном множестве является вычисление
разности в соседних точках. Устойчивость можно обеспечить,
увеличив расстояние между точками, в которых вычисляется разность.
Такой метод известен со времен Ньютона \cite{Tikhonov}.

Одним из практических методов оценки производной в физическом пространстве является свертка измеренного сигнала
со сглаживающим окном (Хэмминга, Винера и др. \cite{Pikalov, Hamming}), с последующим вычислением производной в
виде конечной разности (\ref{fungi}). Сглаженный сигнал находится по формуле

\begin{equation}
\int_{-\infty}^{\infty} f(x') u_{\alpha}(x-x') dx',
\label{eq_conv}
\end{equation}
где $u$ - сглаживающее окно, ширина которого управляется регуляризирующим параметром $\alpha$.

Применяемое на практике усреднение по соседним точкам эквивалентно
сглаживанию прямоугольным окном. Численная оценка погрешности
вычисления производной в физическом пространстве (при
использовании различных сглаживающих окон) будет дана ниже.

\subsection{Приближенное вычисление производной в пространстве Фурье}

Как альтернативу свертке со сглаживающим окном в физическом пространстве рассмотрим умножение на фильтр низких
частот (ФНЧ) в частотной области. Идеальный ФНЧ отсекает все частоты, выше $k_0(\alpha)$. Сигнал, пропущенный
через идеальный ФНЧ, не будет иметь высоких частот, то есть будет сглаженным. После выполнения обратного
преобразования Фурье дифференцирование производится методом конечной разности.

Можно объединить в частотной области обе операции - обеспечения устойчивости путем подавления высоких частот и
вычисления производной путем умножения на $ik$. Алгоритм (\ref{eq_ift}) в условиях шума является неустойчивым,
так как умножение на $ik$ приводит к неограниченному усилению высоких частот. Для обеспечения устойчивости
используется ФНЧ с шириной полосы пропускания, управляемой регуляризирующим параметром (при $\alpha \to 0$,
полоса пропускания стремится к бесконечности и приближенное решение стремится к точному). Форма фильтра может
выбираться достаточно произвольно, исходя из специфики задачи. Ниже будут рассмотрены два фильтра:
дифференцирующий фильтр, представляющий собой произведение $ik$ и идеального ФНЧ, и дифференцирующий гауссов
фильтр - произведение $ik$ и гауссова фильтра.

\subsection{Вейвлет-регуляризация}

При использовании вейвлет-алгоритма вместо свертки сигнала с дублетом выполняется анализ с использованием
вейвлета $\chi$ и последующий синтез с использованием вейвлета $\psi$.

Интегрирование по $a$ на этапе синтеза на практике осуществляется в конечных пределах от $a_{min}$ до $a_{max}$.
В случае высокочастотного шума выбор $a_{min}$ необходимо связать с ожидаемым уровнем шума. В простейшем случае
пределы интегрирования по $a$ выбираются на основании интегральных критериев и не учитывают поведение функции и
особенности шума в различные моменты времени. Преимущество вычисления с помощью вейвлетов состоит в том, что
легко реализовать локальную регуляризацию, когда пределы интегрирования адаптируются под локальные свойства
вейвлет-спектра, т.е. $a_{min}$ зависит от $x$.

\section{Примеры}
\label{part_example}

\begin{figure}   \centering
\includegraphics[width=0.48\textwidth]{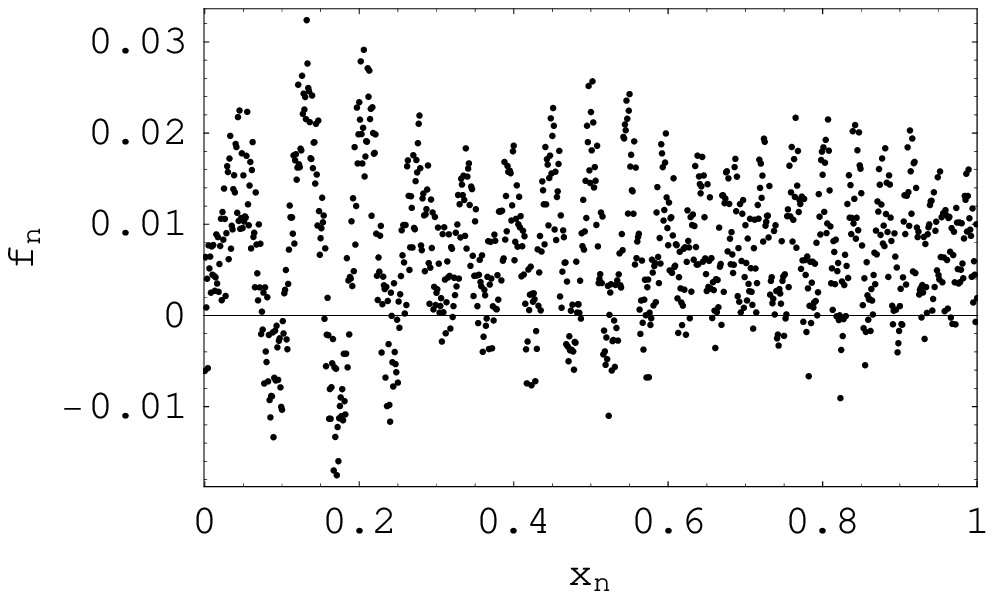}
\includegraphics[width=0.48\textwidth]{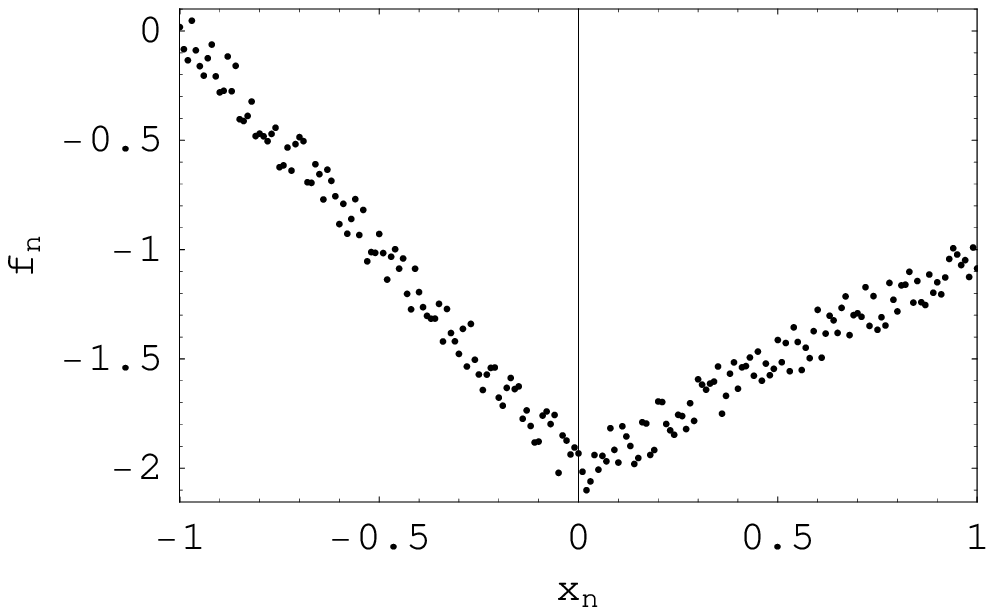}
\includegraphics[width=0.48\textwidth]{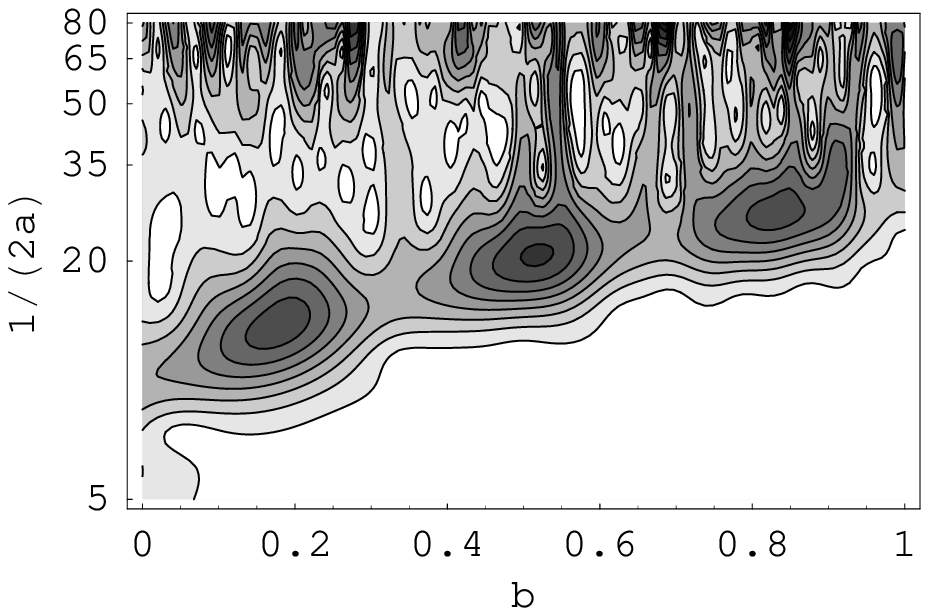}
\includegraphics[width=0.48\textwidth]{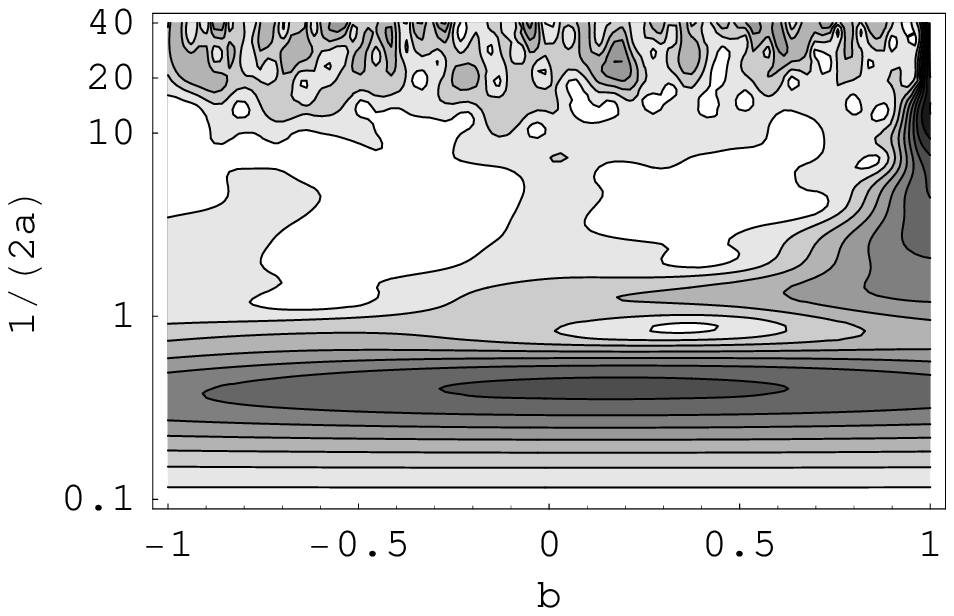}
\caption{Модельные сигналы с аддитивным белым шумом {\it(вверху)} и соответствующие вейвлет-образы {\it
(внизу)}.}\label{fig_sigf}
\end{figure}

Выбор оптимального способа дифференцирования зависит, безусловно, от вида сигнала и характеристик шума. Мы
ограничимся рассмотрением двух характерных примеров. Первый пример представляет сигнал, основная энергия
которого сосредоточена в высокочастотной части спектра (периоды характерных колебаний существенно меньше длины
интервала, на котором задан сигнал). В этом случае можно считать, что сигнал задан на неограниченном отрезке или
является периодическим, что позволяет избежать проблем, связанных с границами области определения. В качестве
второго пример выбрана кусочно-гладкая функция.

Первый пример представляет собой заданный на интервале от 0 до 1 с шагом 0.001 осциллирующий сигнал,
модулированный по частоте и амплитуде, искомая производная которого определяется по формуле
\begin{equation}
g(x)=\sin(2 \pi x (10 x +10))(1-1/2 \cos(6 \pi x)).
\label{funtest}
\end{equation}

Производная второго модельного сигнала определяется выражением
\begin{equation}
g(x)=\left\{
   -2, x\le 0 \atop
   1 , x>0
   \right. .
\label{funtest2}
\end{equation}

Второй сигнал более гладкий и не имеет высокочастотной составляющей за исключением области разрыва производной.
Этот сигнал задан только в интервале от -1 до 1 и ограниченность области определения оказывает прямое влияние на
дифференцирование.

Для каждого из модельных сигналов были сформированы наборы точек $f_n$ по формуле
\begin{equation}
f_n=\int_{x_0}^{x_n} g(x) dx + \xi(\mu), \label{funsig}
\end{equation}
где $\xi(\mu)$ - белый шум с уровнем $\mu*100\%$ от среднего абсолютного значения функции.

Полученные распределения представлены на Рис.~\ref{fig_sigf}. Уровень шума составлял 30\% для первого сигнала
(Рис.~\ref{fig_sigf}, слева) и 10\% для кусочно-гладкого примера (Рис.~\ref{fig_sigf},~справа). В
высокочастотной части соответствующих вейвлет-образов (Рис.~\ref{fig_sigf},~внизу) наблюдаются нерегулярные
структуры, вызванные добавлением шума.

Массивы значений $f_n$ использовались для численного вычисления производной $g_n$ различными алгоритмами и
сравнения с аналитическими значениями производной $g(x_n)$. Среднеквадратичная ошибка (СКО) дифференцирования
определялась по формуле \cite{Pikalov}
\begin{equation}
\sigma=\sqrt{\frac{\sum_n(g_n - g(x_n))^2}{\sum_n {g(x_n)}^2}},\label{funsigma}
\end{equation}
где $g(x_n)$ - значения производной модельного сигнала в точках $x_n$, $g_n$ - результат численного
дифференцирования.

\begin{figure}   \centering
\includegraphics[width=0.48\textwidth]{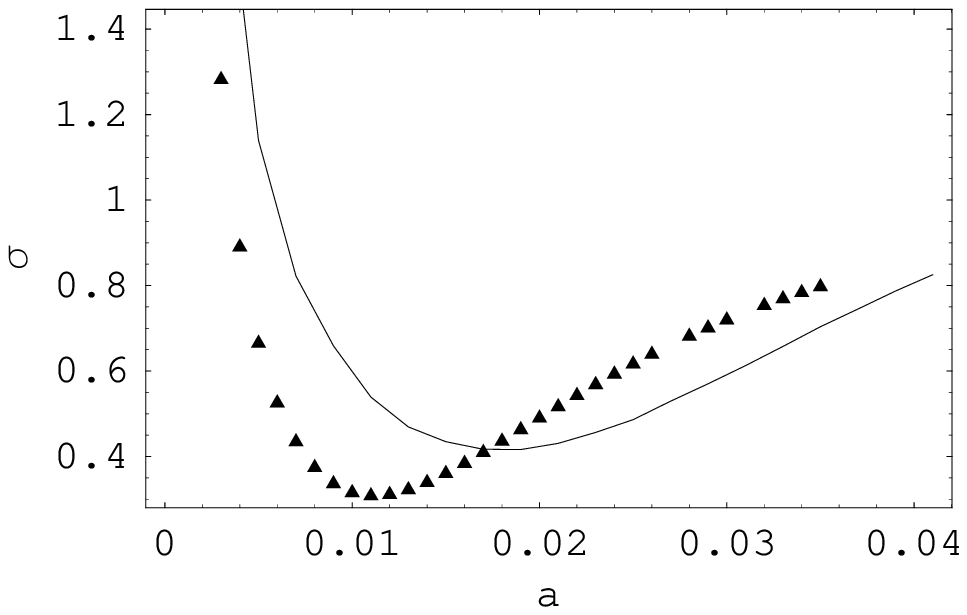}
\includegraphics[width=0.48\textwidth]{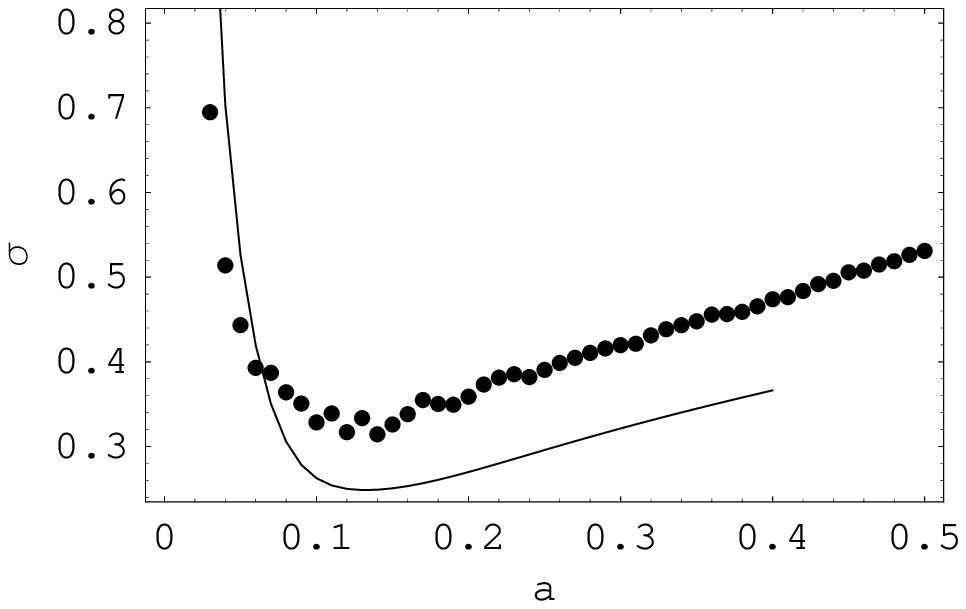}
\includegraphics[width=0.48\textwidth]{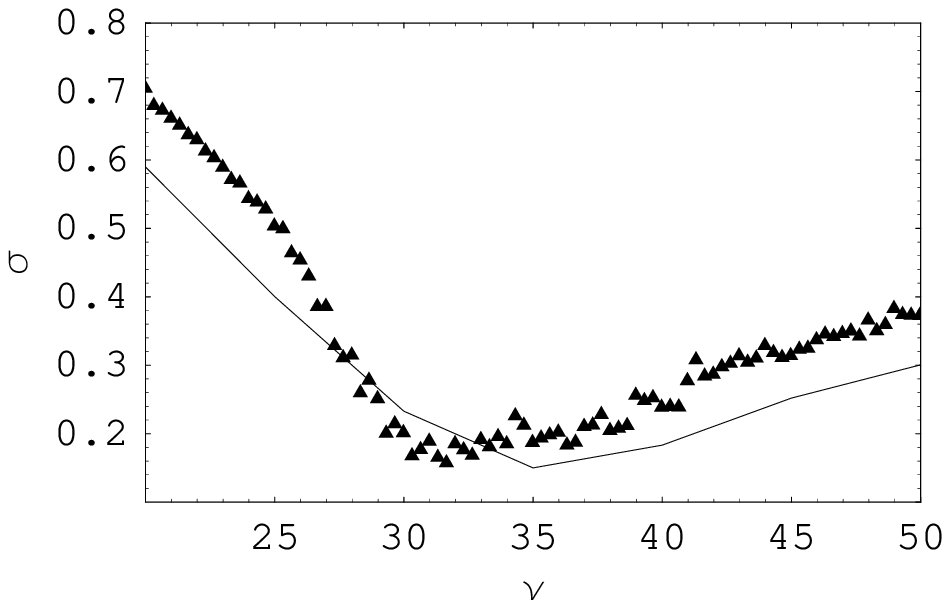}
\includegraphics[width=0.48\textwidth]{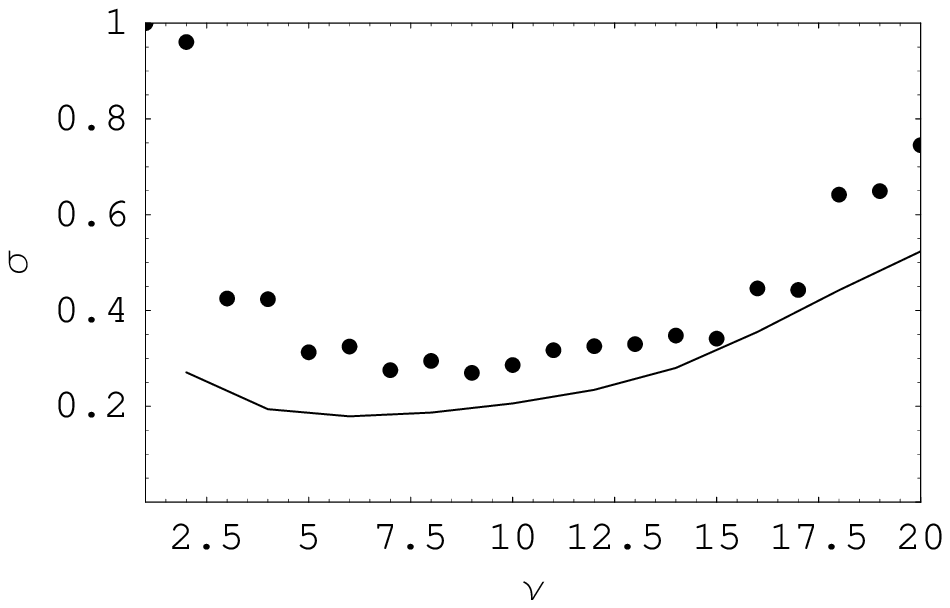}

\caption{Зависимость среднеквадратичной ошибки дифференцирования (пример~1 - слева, пример~2 - справа) от ширины
окна в физическом пространстве: прямоугольное окно - линия, гауссово окно - точки {\it(вверху)}; в частотном
пространстве: прямоугольный ФНЧ - точки, вейвлет-дифференцирование - линия {\it(внизу)}. }\label{fig_sigma}
\end{figure}

\begin{table}
\centering
\begin{tabular}{|l|c|c|}
\hline
Метод & \multicolumn{2}{c}{Минимальное значение $\sigma$}\vline\\
\cline{2-3} & Пример 1 & Пример~2 \\ \hline
Прямоугольное окно               & 0.42 & 0.31 \\
Гауссово окно                    & 0.31 & 0.25 \\
\hline
Фурье (ФНЧ)+кон.разн.            & 0.16 & 0.8  \\
Фурье гаусс + кон.разн.          & 0.16 & 0.7  \\
Дифференцирующий фильтр          & 0.18 & 0.27 \\
Диф.гаусс фильтр                 & 0.3  & 0.32 \\
\hline
Вейвлет-регуляризация (Морле)    & 0.15 & 0.2  \\
Вейвлет $a_{min}=(20*x+c)^{-1}$  & 0.12 & -    \\
\hline
\end{tabular}
\caption{Сравнение эффективности фильтрации шума при вычислении производной.}\label{table1}
\end{table}

В данной работе сравнивались следующие алгоритмы:

\emph{Прямоугольное окно в физическом пространстве + конечная разность.} Сигнал усреднялся по соседним точкам и
затем применялся метод конечной разности для оценки производной. Оптимальное значение ширины окна,
минимизирующее $\sigma$, определялось путем варьирования количества точек, по которым проводится усреднение, и
последующим сравнением с аналитическим решением. Оптимальное значение ширины окна имеет значение 0.018 для
первого сигнала и 0.14 - для второго сигнала. Минимальные значения $\sigma$ приведены в Таблице~\ref{table1}.
Графически зависимость $\sigma$ от ширины окна показана на Рис.~\ref{fig_sigma}.

\emph{Гауссово окно в физическом пространстве + конечная разность.} Аналогичным образом определось оптимальное
значение ширины гауссова окна (на уровне половины от максимального значения). Оптимальное значение ширины окна
равно 0.011 для первого сигнала и 0.31 - для второго сигнала. Минимальные значения $\sigma$ - в
Таблице~\ref{table1}. Сравнивая графики зависимости $\sigma$ от ширины окна (Рис.~\ref{fig_sigma}), можно
видеть, что гауссово окно дает более точное восстановление производной, чем прямоугольное.

\emph{ФНЧ в частотном пространстве + конечная разность.} Оптимальное значение частоты отсечки $k_0$,
минимизирующее $\sigma$, определялось следующим образом: вычислялось FFT сигнала, отбрасыались коэффициенты ряда
Фурье, соответствующие частотам $k>k_0$, а затем выполнялось обратное FFT и применялся метод конечной разности.
Зависимость $\sigma$ от $k_0$ представлена на Рис.\ref{fig_sigma}. Графически зависимость $\sigma$ от $k_0$
показана на Рис.~\ref{fig_sigma}. Минимальное значение $\sigma$, достигаемое при использовании идеального
фильтра низких частот, - в Таблице~\ref{table1}.

\emph{Гауссов фильтр в частотном пространстве + конечная разность.} Фильтр получается заменой прямоугольного ФНЧ
на гауссов. Зависимость $\sigma$ от ширины фильтра в частотном пространстве находилась следующим образом:
вычислось FFT сигнала, умножались коэффициенты ряда Фурье на гауссов фильтр, а затем выполнялось обратное FFT и
применялся метод конечной разности. Оптимальное значение ширины фильтра равно 32 для первого примера и 1 - для
второго примера. Минимальные значение $\sigma$ - в Таблице~\ref{table1}.

\emph{Дифференцирующий фильтр в частотном пространстве.} Умножение идеального ФНЧ на $ik$ дает фильтр
специальной формы, осуществляющий в частотной области вычисление производной одновременно со сглаживанием. Такой
фильтр, а также его ближайшие родственники, используются в томографии для вычисления обратного преобразования
Радона \cite{Pikalov,Levin}. Оптимальное значение частоты отсечки $k_0$ равно 32 для первого примера  и 8 - для
второго примера. Минимальные значение $\sigma$ - в Таблице~\ref{table1}.

\emph{Дифференцирующий гауссов фильтр в частотном пространстве.} Модифицируем предыдущий фильтр, заменив
идеальный ФНЧ на гауссов фильтр. Умножив в частотном пространстве гауссов фильтр на $ik$, получим фильтр
$ik\exp(-{k}^2)$, импульсная характеристика которого равна первой производной от гауссова окна. Варьируя ширину
фильтра в частотной области, найдем значение параметра масштаба, минимизирующее $\sigma$. Оптимальное значение
ширины фильтра равно 37 для первого примера и 10 - для второго примера. Минимальные значения $\sigma$
представлены в Таблице~\ref{table1}.

\emph{Вейвлет Морле в физическом пространстве.} Разложим сигнал по масштабам, используя производную от вейвлета
Морле в качестве анализирующего вейвлета. Диапазон масштабов зададим от 0.1 до 0.006, за пределами которого
сигнал практически отсутствует. Затем для каждого масштаба вычислим свертку полученных вейвлет-коэффициентов с
вейвлетом Морле и проинтегрируем по масштабам от $a_{max}=0.1$ до $a_{min}$. Варьируя $a_{min}$, найдем
соответствующие значения $\sigma$. Зависимость $\sigma$ от $a_{min}$ (переведенного в частотный аналог по
формуле $\nu=1/(2 a_{min})$) представлена на Рис.~\ref{fig_sigma}. Оптимальное значение $\nu=35$ для первого
примера и 7 - для второго примера. Минимальное значение $\sigma$ - в Таблице~\ref{table1}. Анализ вейвлет-образа
квазигармонического сигнала (Рис.\ref{fig_sigf}, слева) показывает, что существенная часть информации
сосредоточена в трех компактных областях, расположенных вдоль наклонной прямой. Это дает возможность задать
минимальный масштаб в виде функции $a_{min}=(10*x+c)^{-1}$. Тогда перебор значений $c$ дополнительно
минимизирует $\sigma$. В этом случае для значения $c=42$ получается результат $\sigma=0.12$, что заметно лучше,
чем результат, полученный с использованием преобразования Фурье. Для кусочно-гладкого сигнала (Пример~2)
дополнительная оптимизация не проводилась, так как его вейвлет-образ (Рис.\ref{fig_sigf}, справа) существенно
искажен влиянием шума.

\begin{figure}
\centering
\includegraphics[width=0.48\textwidth]{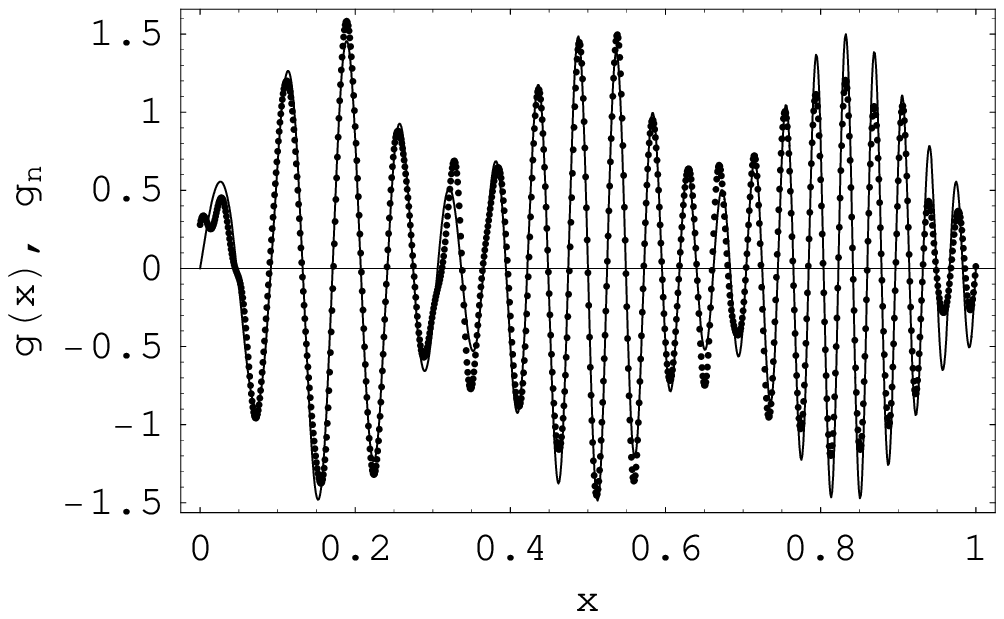}
\includegraphics[width=0.48\textwidth]{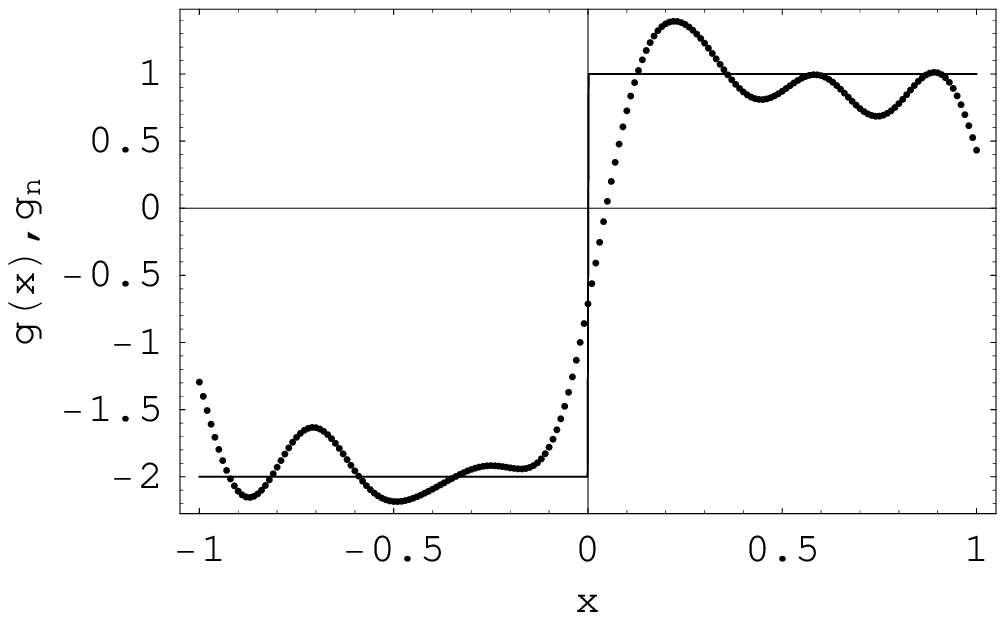}
\includegraphics[width=0.48\textwidth]{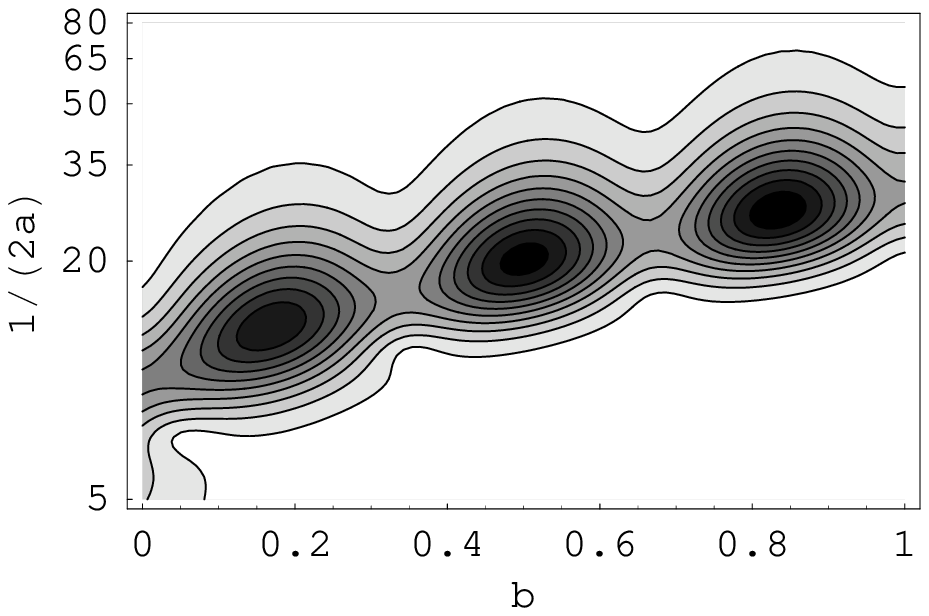}
\includegraphics[width=0.48\textwidth]{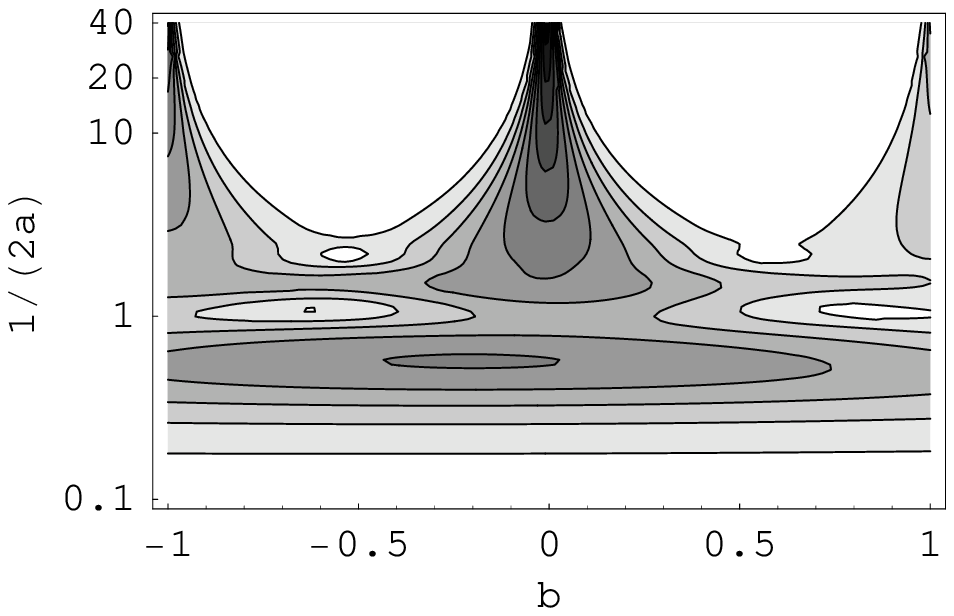}
\caption{Производные модельных сигналов, вычисленные аналитические - линии; численная оценка с использованием
вейвлета Морле при оптимальном значении $a_{min}$ - точки {\it(вверху)}. Вейвлет-образы производных модельных
сигналов {\it(внизу)}.}
\label{fig_diff}
\end{figure}

На Рис.\ref{fig_diff} представлены графики аналитических производных модельных сигналов и результаты численного
дифференцирования методом на основе вейвлета Морле при оптимальном значении $a_{min}$ (без использования
дополнительной оптимизации). Вейвлет-образы производных сигналов, приведенные на Рис.~\ref{fig_diff},внизу,
показывают распределения спектральных свойств в заданных интервалах.

\section{Обсуждение}

Алгоритм c использованием вейвлет-преобразования позволяет проводить устойчивое дифференцирование в условиях
шума. Локальность базисных функций позволяет более точно учитывать свойства сигнала по сравнению с методами
фильтрации в частотном пространстве.

Особый интерес предложенные методы могут представлять при решении обратных задач (в медицине, астрономии,
гидродинамике и т.д.). Например, интегральное уравнение Абеля, к которому сводится задача осесимметричной
томографии, имеет решение \cite{Levin}
\begin{equation}
g(r)=-\frac{1}{\pi}\int_{r}^{\infty} \frac{dp}{\sqrt{p^2-r^2}} \frac{\partial f(p)}{\partial p},
\label{eq_abel}
\end{equation}
где $f(p)$ - измеренные с некоторой погрешностью проекционные данные.

В общем случае решение двумерной задачи томографии сводится к обратному преобразованию Радона, которое также
может быть выражено через первую производную измеренного сигнала \cite{Pikalov}
\begin{equation}
g(x,y)=-\frac{1}{2\pi^2}\int_{0}^{\pi} d\phi \int_{-\infty}^{\infty} \frac{dp}{p-x\sin\phi+y\cos\phi}
\frac{\partial f(p,\phi)}{\partial p}, \label{eq_irt}
\end{equation}
где $f(p,\phi)$ - измеренные с некоторой погрешностью проекционные данные.

В работе \cite{frick_etal_98} приведен ряд примеров, показывающих, что метод позволяет эффективно подавить шумы,
возникающие в спектре из-за пробелов, и повысить точность восстановления спектра исходного сигнала. Этот
алгоритм легко переносится и на рассмариваемую задачу о вычислении производной и состоит в этом случае в
следующем. На первом шаге с помощью метода "дырявых" вейвлетов вычисляется вейвлет-образ $W_{a,b}$ исходной
функции $\tilde f(x)$, а на втором шаге - по формуле (\ref{iwt}) восстанавливается искомая производная
$g(x)=f'(x)$. При этом на втором шаге используются {\it полные} вейвлет-функции $\chi(x)$.

\vspace{5mm} {\bf Благодарности.} Работа выполнена при финансовой поддержке гранта РФФИ-ННИО № 03-02-04031, РФФИ
№ 03-02-16384 и Научно-образовательного центра (грант PE-009-0). РС также благодарен Уральскому отделению РАН
(грант молодым ученым).

\end{document}